\documentclass[english,12pt]{article}
\usepackage{array}
\usepackage{graphicx}
\usepackage{amssymb}
\usepackage{amsmath}
\usepackage{multirow}
\usepackage{prettyref}
\usepackage{babel}
\usepackage{units}
\usepackage[latin1]{inputenc}
\usepackage{amsfonts}
\usepackage{amssymb}
\usepackage{babel}
\usepackage{cite}
\def\@fmsl@sh#1#2#3{\m@th\ooalign{$\hfil#1\mkern#2/\hfil$\crcr$#1#3$}}
 \def\eq#1\en{\begin{equation}#1\end{equation}}
\def\s[#1,#2]{[#1\stackrel{\star}{,}#2]}
\def\sx[#1,#2]{[#1\stackrel{\star_{x}}{,}#2]}

\newcommand{\nc}{\newcommand}
\nc{\beq}{\begin{equation}}
\nc{\eeq}{\end{equation}}
\nc{\beqa}{\begin{eqnarray}}
\nc{\eeqa}{\end{eqnarray}}

\def\bc{\begin{center}}
\def\ec{\end{center}}

\def\gsim{\mathrel{\mathpalette\atversim>}}

\def\bc{\begin{center}}
\def\ec{\end{center}}

\def\gsim{\mathrel{\rlap{\lower4pt\hbox{\hskip1pt$\sim$}}

    \raise1pt\hbox{$>$}}}       

\def\gsim{\mathrel{\rlap{\lower4pt\hbox{\hskip1pt$\sim$}}
    \raise1pt\hbox{$>$}}}       

\begin{document}
\makeatletter
\def\fmslash{\@ifnextchar[{\fmsl@sh}{\fmsl@sh[0mu]}}
\def\fmsl@sh[#1]#2{%
  \mathchoice
    {\@fmsl@sh\displaystyle{#1}{#2}}%
    {\@fmsl@sh\textstyle{#1}{#2}}%
    {\@fmsl@sh\scriptstyle{#1}{#2}}%
    {\@fmsl@sh\scriptscriptstyle{#1}{#2}}}
\def\@fmsl@sh#1#2#3{\m@th\ooalign{$\hfil#1\mkern#2/\hfil$\crcr$#1#3$}}
\makeatother

\thispagestyle{empty}
\begin{titlepage}
\boldmath
\begin{center}
  \Large {\bf The Lightest of Black Holes}
    \end{center}
\unboldmath
\vspace{0.2cm}
\begin{center}
{  {\large Xavier Calmet}\footnote{x.calmet@sussex.ac.uk}}
 \end{center}
\begin{center}
{\sl Physics $\&$ Astronomy, 
University of Sussex,   Falmer, Brighton, BN1 9QH, United Kingdom 
}
\end{center}
\vspace{5cm}
\begin{abstract}
\noindent
In this paper we consider general relativity in the large $N$ limit, where $N$ stands for the number of particles in the model. Studying the resummed graviton propagator, we propose to interpret its complex poles as  black hole precursors.  Our main result is the calculation of the mass and width of the lightest of black holes. We show that the values of the masses of black hole precursors depend on the number of fields in the theory. Their masses can be lowered down to the TeV region by increasing the number of fields in a hidden sector that only interacts gravitationally with the standard model.

\end{abstract}  
\end{titlepage}



\newpage

Nearly a hundred years after Einstein proposed his theory of gravitation, it is still unknown how to  quantize general relativity. We nevertheless have a few hints of what to expect from a theory of quantum gravity. For example, thought experiments lead to the conclusion that a unification of quantum mechanics and general relativity should incorporate the notion of a minimal length \cite{Mead:1964zz,Garay:1994en,Padmanabhan:1987au,Calmet:2004mp}.  A minimal length could be a sign of non-local interactions at the Planck scale. Another interesting feature of a quantum mechanical description of gravitation is the generalized uncertainty principle. Amati, Ciafaloni and Veneziano (ACV)  have shown \cite{Amati:1988tn,Amati:1992zb,Amati:1993tb,Amati:2007ak} by studying the scattering of massless strings (e.g. two gravitons) in the trans-Planckian regime that the usual uncertainty principle of quantum mechanics $\Delta x \Delta p > \hbar$ generalizes to  $\Delta x \Delta p > \hbar +\alpha^\prime \Delta p^2$ where $\alpha^\prime=G_N/g^2$, $G_N$ is Newton's constant and $g$ is the string loop expansion parameter. While this specific relation was derived in string theory similar ones appear in different models of quantum gravity, see e.g. \cite{Hossenfelder:2012jw} for a recent review.

The generalized uncertainty principle finds a natural interpretation in the scattering at high energies of two particles. When two particles collide at energies around the Planck scale a quantum black hole will form (by quantum black hole we mean a Planck size black hole with a mass of the order of the Planck mass). As the center of mass energy increases, so does the mass of the black hole. The black hole becomes larger and is, in a sense, more extended or non-local. Thus increasing the center of mass energy of the scattering experiment does not allow to resolve finer details as the $\Delta x$ probed by the scattering experiment increases with energy. There is thus a minimal length which is given by the Planck length. ACV have suggested that by studying the scattering of massless strings in the eikonal approximation, they can identify  the formation of an object that looks like a black hole precursor.  Obviously, string theory incorporates the notion of non-locality since the fundamental building blocks of these type of models are strings and d-branes which are extended objects. But, we shall see that these ideas can resurface in linearized general relativity coupled to quantum field theory as well.

In this paper, we  study another indication that a unification of quantum mechanics and general relativity must lead to non-local effects. Our main result is a calculation of the mass and width of the lightest black hole. These black holes lead to tiny acausal effects at energy scales comparable to the Planck scale. We show that the mass of the black hole precursors is dependent on the number of fields in the theory.

Recently, there has been  a renewed interest in the gravitational scattering of fields and the question of whether perturbative unitarity could be violated below the Planck scale \cite{Han:2004wt,Atkins:2010eq,Atkins:2010re,Atkins:2010yg,Antoniadis:2011bi,Xianyu:2013rya,Ren:2014sya,Aydemir:2012nz,Calmet:2013hia}. By studying the two to two elastic gravitational scattering of fields, it has been argued that perturbative unitarity is violated at an energy scale $E \sim \bar M_P/\sqrt{N}$  \cite{Han:2004wt}, where $N$ is loosely speaking the number of fields in the model and $\bar M_P$ the reduced Planck mass.  However, it has been shown in  \cite{Aydemir:2012nz} that perturbative unitarity is restored by resumming an infinite series of matter loops on a graviton line (see Fig. 1) in the large $N$ limit, where $N$ stands for the number of fields in the model, while keeping $N G_N$ small.
\begin{figure}[htp]
\centering
\includegraphics[ width=5in]{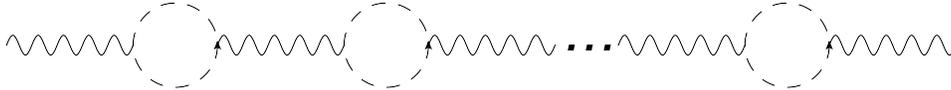}
\caption{Resummation of the gravitaton propagator.}\label{Fig1}
\end{figure}
This large $N$ resummation leads to resummed graviton propagator given by
\begin{eqnarray} \label{resprop}
i D^{\alpha \beta,\mu\nu}(q^2)=\frac{i \left (L^{\alpha \mu}L^{\beta \nu}+L^{\alpha \nu}L^{\beta \mu}-L^{\alpha \beta}L^{\mu \nu}\right)}{2q^2\left (1 - \frac{N G_N q^2}{120 \pi} \log \left (-\frac{q^2}{\mu^2} \right) \right)}
\end{eqnarray}
with $L^{\mu\nu}(q)=\eta^{\mu\nu}-q^\mu q^\nu /q^2$, $N=N_s +3 N_f +12 N_V$ where $N_s$, $N_f$ and $N_V$ are respectively the number of real scalar fields, fermions and spin 1 fields in the model. This mechanism was dubbed self-healing by the authors of \cite{Aydemir:2012nz}. While \cite{Aydemir:2012nz} emphasized the fact that perturbative unitarity is restored by the resummation, the authors of \cite{Han:2004wt} who had studied the same phenomenon before had pointed out that the denominator of this resummed propagator has a pair of complex poles which lead to acausal effects (see also \cite{Tomboulis:1977jk,Tomboulis:1980bs} for earlier work in the same direction and where essentially the same conclusion was reached). These acausal effects should become appreciable at energies near $(G_N N)^{-1/2}$. Unitarity is restored but at the price of non-causality.

We propose to interpret these complex poles as Planck size black hole precursors or quantum black holes. This enables us to calculate the mass and the width of the lightest black hole. This pair of complex poles which appears at an energy of about $(G_N N)^{-1/2}$ is a sign of strong gravitational dynamics. It is thus natural to think that this is the energy scale at which black holes start to form. Note that our interpretation is not controversial, one expects black holes to have a lifetime of order their Schwarzschild radius and thus to be described by propagators of the type $(s-M_{BH}^2 +i M_P^2 )^{-1}$ \cite{Amati:2007ak}. 
Let us now calculate the poles of the resummed propagator (\ref{resprop}).  We find
\begin{eqnarray}
q^2_{1}&=0,\\ \nonumber
q^2_{2}&=& \frac{1}{G_N N} \frac{120 \pi}{ W\left (\frac{-120 \pi M_P^2}{\mu^2 N} \right)}, \\ \nonumber
q^2_{3}&=&(q^2_{2})^*,
\end{eqnarray}
where $W(x)$ is the Lambert W-function. It is easy to see that for $\mu \sim M_P$, $q_{2/3} \sim (G_N N)^{-1/2}$ as mentioned previously. The resummed propagator has  three poles, one at $q^2=0$ which corresponds to the usual massless graviton and a pair of complex poles $q^2_{2,3}$.
 In the standard model of particle physics, one has $N_s = 4$, $N_f = 45$, and $N_V = 12$. We thus find $N=283$ and the pair of complex poles at $(7 -  3 i)\times 10^{18}$ GeV and  $(7 + 3 i)\times 10^{18}$ GeV. The first of these pair of poles corresponds to an object with mass $7\times 10^{18}$ GeV with a width $\Gamma$ of $6\times 10^{18}$ GeV. In our interpretation, these are the mass and width of the lightest of black holes assuming that  the standard model of particle physics is valid up to the Planck scale\footnote{Note that in \cite{Aydemir:2012nz}, it was argued that one could identify the $\sigma$-meson as the pole of a resummed scattering amplitude in the large $N$ limit of chiral perturbation theory. This resummed amplitude is an example of self-healing in chiral perturbation theory.  In low energy QCD, the position of the pole does correspond to the correct value of the mass and width of the $\sigma$-meson.}. It is a quantum black hole with a mass just above the reduced Planck scale ($2.435\times10^{18}$ GeV) and a lifetime given by $1/\Gamma$. Obviously, these estimates depend on the renormalization scale. Since the only scale in the problem is the reduced Planck scale, here we have taken $\mu$ of the order of the reduced Planck scale. We have checked that our predictions are not numerically very sensitive to small changes of the renormalization scale. Note that we have used the definition for the mass and width introduced in \cite{Bhattacharya:1991gr}, namely we identify the mass and width of the black hole precursor with the position of pole in the resummed propagator: $p_0^2=(m-i\Gamma/2)^2$. The second complex pole  at $(7 + 3 i)\times 10^{18}$ GeV leads to the acausal effects.

Since black holes are extended objects with a radius $R_S= 2 G_N M/c^2$, it is not surprising that they lead to non-local effects. It has been shown in \cite{Donoghue:2014yha} that the momentum space equivalent of the non-local term in the resummed propagator is of the type 
\begin{eqnarray}
S= \int d^4 x \sqrt{g} \left [ R \log \left ( \frac{\Box }{\mu^2} \right ) R \right ].
\end{eqnarray}
Furthermore, it has been argued by Wald in \cite{Wald:1995yp} that when the space-time metric is treated as a quantum field, there should be fluctuations in the local light cone structure which could be large at the Planck scale. These fluctuations imply that the causal relationships between events may not be well defined and that there is a nonzero probability for acausal propagation at energies around the Planck scale. The Planckian black hole we are studying here is the black hole for which  quantum gravitational effects are the most important of all, it is thus not very surprising that it leads to acausal effect according to Wald's argument. Note that acausal effects of this type have been discussed in the framework of the Lee Wick formalism \cite{Lee:1969fy,Lee:1970iw} (see also \cite{Grinstein:2008bg} for more recent work in that direction).

With our  interpretation in mind,  a consistent and beautiful picture emerges. Self-healing in the case of gravitational interactions implies unitarization of quantum amplitudes via quantum black holes. As the center of mass energy increases so does the mass of the black hole and it becomes more and more classical. This is nothing but classicalization \cite{Calmet:2010cb,Dvali:2010jz}.  Furthermore, one expects as well a modification of the uncertainty relation of the type:
\begin{eqnarray}
\Delta x \Delta p > \hbar +\alpha f(\Delta p^2),
\end{eqnarray}
where the parameter $\alpha$ is positive. 
 As mentioned before, as we increase the center of mass energy, so does the mass of the black hole in the pole of the resummed propagator. The black hole becomes larger and the magnitude of the nonlocal effects increases. Thus, as in the case of ACV,  increasing the center of mass energy of the scattering experiment does not allow to resolve shorter distances as the $\Delta x$ probed by the scattering experiment increases with the center of mass energy. Since we cannot trust our calculation in the trans-Planckian regime we cannot calculate the function $f(\Delta p^2)$ in contrast to what ACV had done using the eikonal approximation in string theory.  Note however that in the large $N$ limit we are using, loops of gravitons are always suppressed compared to matter loops. We are in a situation where a full knowledge of quantum gravity may not be necessary. A similar point has been made by 't Hooft \cite{tHooft:2011aa} in a different context. 

It is worth mentioning that a potential non-minimal coupling $\xi$ of the scalar fields  to the Ricci scalar plays no role in the resummed propagator (\ref{resprop}). A non-minimal coupling of scalars to the Ricci scalar does not affect the mass of black hole precursors. This is consistent with the results obtained in  \cite{Calmet:2013hia} where it was shown that the large $\xi N$ limit leads to a resummed graviton propagator which does not have a pole. In other words, models such as Higgs inflation which rely on a non-minimal coupling of the Higgs boson to curvature are perfectly valid and there is no sign of strong dynamics below the Planck scale. 

Our approach leads to a new mechanism to lower the energy scale of quantum gravity below the traditional Planck scale or some $10^{19}$ GeV. For a large number of fields, e.g. $N_s=10^{33}$ and  for example $\mu=1$ TeV, the Planckian black hole precursor has a mass of $3.7$ TeV with a width of $2.8$ TeV.  This new mechanism demonstrates that, as observed previously in \cite{Calmet:2008tn, Dvali:2008ec,Dvali:2012uq}, a large number of particles in a hidden sector that only interact gravitationally with the standard model leads to a lowering of the effective Planck mass. While in \cite{Calmet:2008tn}, a renormalization group of Newton's constant was used to describe this effect, here we do not have to rely on such a technique which can potentially lead to ambiguous results \cite{Anber:2011ut}. For some $10^{33}$ fields in the hidden sector, the Planck mass is around a TeV and Planck mass quantum black holes  \cite{Meade:2007sz,Calmet:2008dg} could be produced at the Large Hadron collider.  Current searches for quantum black holes at the LHC thus imply a bound on the number of particles that couple to the graviton. The non-observation of quantum black holes at the LHC thus far \cite{Aad:2013gma} implies that the Planck scale is above 5 TeV. This translates into a bound on the number of fields gravitationally coupled to the standard model: $N<10^{33}$.

In this brief paper we have calculated the mass and width of the lightest of black holes. We have shown that the values of these parameters are dependent on the number of fields in the theory. In the case of the standard model,  these results are consistent with expectations: we find that both the mass and the width of the lightest black hole are of the order of the reduced Planck scale. Interpreting the poles of the resummed graviton propagator in the large $N$ limit leads to a beautiful insight into the unification of quantum mechanics and general relativity. Non-causality seems to be a feature of such a unification in the form of quantum black holes and it may be a sign that quantum gravity is made finite by a mechanism of the Lee Wick type. The self-healing mechanism and the classicalization mechanism appear to be necessary ingredients of quantum gravity and the generalized uncertainty principle a necessary consequence of these mechanisms. Finally, the effective Planck mass depends on the number of fields in the theory. The mechanism can be used to build models with low scale quantum black holes. Using recent data from the Large Hadron Collider at CERN, we have derived a bound on the number of fields which belong to a hidden sector that only interacts gravitationally with the standard model.

{\it Acknowledgments:}
This work is supported in part  by the Science and Technology Facilities Council (grant number  ST/J000477/1).


\bigskip{}

\end{document}